# Sponges and Engines
## An introduction to Keccak and Keyak

Jos Wetzels, Wouter Bokslag

`a.l.g.m.wetzels@student.utwente.nl`

`w.bokslag@student.tue.nl`

**Abstract.** In this document we present an introductory overview of the algorithms and design components underlying the *Keccac* cryptographic primitive and the *Keyak* encryption scheme for authenticated (session-supporting) encryption. This document aims to familiarize readers with the basic principles of *authenticated encryption*, the *Sponge* and *Duplex* constructions (*full-state*, *keyed* as well as regular versions), the permutation functions underlying *Keccak* and *Keyak* as well as *Keyak v2*'s Motorist mode of operation.

## 1 Introduction

This document presents an overview of the algorithms and design components underlying the *Keccac* [18] cryptographic primitive (a subset of which has been standardized by NIST as the SHA-3 cryptographic hash function [19]) and the *Keyak v2* [24] (hereafter referred to simply as *Keyak*) encryption scheme for authenticated (session-supporting) encryption.

The goal of this document is to provide readers with an overview of the goals and need for *Authenticated Encryption*, especially within the context of the *CAESAR* competition [16], and the permutations and constructions underlying the *Keccak* and *Keyak* families. This document does not seek to compile a state-of-the-art overview of either permutation-based encryption or *Sponge* (and related) constructions nor does it present original research or seek to improve upon the current state-of-the-art. Nor is this document intended to be an implementation reference and as such implementers are referred to the original specification documents for *Keccak* [18] and *Keyak* [24].

Instead it seeks to gather and present available documentation on these matters in a single, accessible introductory document (omitting design considerations and justifications, security proofs and formalizations for the sake of brevity and clarity) aimed at students and industry professionals.

## 2 Authenticated Encryption

Authenticated Encryption (AE) or Authenticated Encryption with Associated Data (AEAD) is a cryptographic mode of operation [1] providing confidentiality, integrity and authenticity assurances on data where decryption is combined in single step with integrity validation. Data authentication is of importance in scenarios where a *Man-in-the-Middle* (MitM) attacker can make arbitrary modifications to the senders' ciphertext before it is received by the recipient. This can lead to a whole host of problems ranging from (in the simplest case) data corruption to bit-flipping attacks [2] and padding oracle attacks [3]. Cryptographic schemes which only guarantee confidentiality are at risk of such attacks [4].

There are two categories of modes of operation providing confidentiality and authenticity: AE and AEAD. The difference between the two is that the latter allows for the authentication of data separate from the plaintext (known as *Additional Authenticated Data* (AAD)) which does not require confidentiality and is only authenticated and not encrypted. This serves to prevent modifications to metadata (eg. ip address and port in network data or e-mail headers in email data) by a MitM attacker. In this document when we refer to AE we refer to both categories unless explicitly stated otherwise.

The need for AE arose out of observations that combining block cipher confidentiality modes with block cipher authentication modes proved easy to get wrong in practice [5]. Combining secure ciphers with secure MACs could still result in insecure authenticated encryption schemes as shown by practical attacks against eg. SSL/TLS [6,7].

An example AE mode API would look as follows:

$$(c, a) = \text{Encrypt}(p, k, \text{optional } h)$$
$$p' = \text{Decrypt}(c, k, a, \text{optional } h)$$

For ciphertext $c$, authentication tag $a$, plaintext $p$, secret key $k$, optional header $h$ and decrypted ciphertext $p'$.

The optional *header* is the *Additional Authenticated Data* (AAD) discussed above. The *authentication tag* consists of a *Message Authentication Code* (MAC). Decryption outputs either the *plaintext* or an error if the provided *authentication tag* does not match the provided *ciphertext* or optional *header*.

AE comes in the form of either specialized block cipher modes of operation (eg. OCB, EAX, CCM, GCM, etc.) [8] or a general construction (not limited to block ciphers) combining an encryption scheme and a MAC provided that the encryption scheme is secure under a *Chosen Plaintext Attack* (CPA) and the MAC is unforgeable under a *Chosen Message Attack* (CMA). Three widespread compositions of these primitives are listed by Krawczyk [9] as follows:

- *Encrypt-then-Authenticate (EtA)*: Here the plaintext is first encrypted after which a MAC of the ciphertext is generated as the *authentication tag*. The result is the pair $(ciphertext, tag = MAC(ciphertext))$.

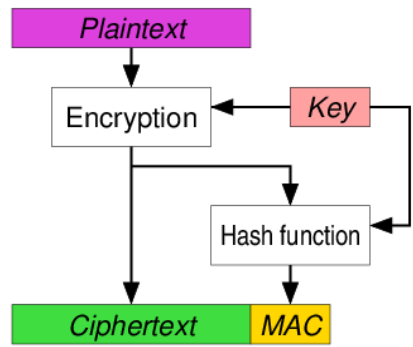

**Fig. 1**: *Encrypt-then-Authenticate* [10]

- *Encrypt-and-Authenticate (AaM)*: Here a MAC of the plaintext is produced and the plaintext is encrypted separately resulting in the pair $(ciphertext, tag = MAC(plaintext))$.

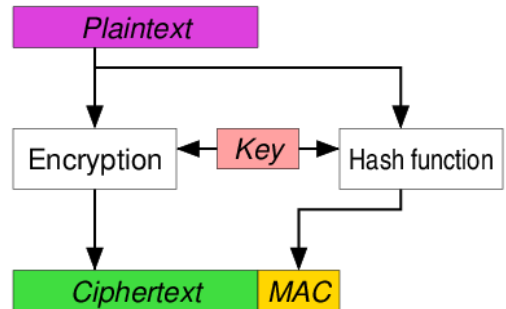

**Fig. 2**: *Encrypt-and-Authenticate* [10]

- *Authenticate-then-Encrypt (AtE)*: Here a MAC of the plaintext is generated and concatenated with the plaintext with the result being encrypted resulting in $(ciphertext = E(plaintext|MAC(plaintext), K)$.

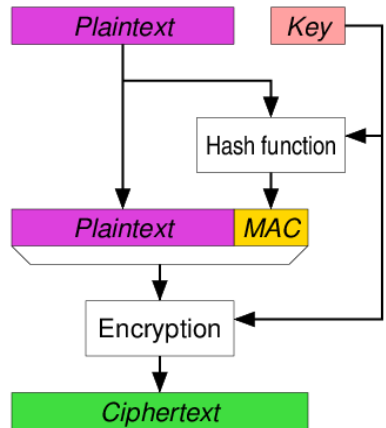

**Fig. 3**: *Authenticate-then-Encrypt* [10]

Note that in the above figures 1 to 3 the same key is used for encryption and authentication while it is recommended [11] to use two different secret keys for these different purposes.

The *Encrypt-then-Authenticate* approach has been shown [9] to be secure (provided the underlying primitives meet the appropriate security requirements). While *Encrypt-and-Authenticate* and *Authenticate-then-Encrypt* have not been shown to be secure by themselves some *implementations* have been shown to be secure. On the other hand while Krawczyk initially reported the *AtE* scheme used in SSL/TLS to be secure he revised his results [12] and it was found that using SSL/TLS with a block cipher in CBC mode was insecure due to the way in which the plaintext was encoded and padded [13].

Given these implementation troubles NIST specified [14,15] two block cipher modes of operation (namely CCM and GCM) offering AE functionality (or rather, AEAD functionality to be more precise). It is important to note that when using randomized IVs appended to the ciphertext one should authenticate the $(IV, ciphertext)$ pair.

### 2.1 CAESAR

CAESAR [16] (Competition for Authenticated Encryption: Security, Applicability, and Robustness) is a project to "identify a portfolio of authenticated ciphers that (1) offer advantages over AES-GCM and (2) are suitable for widespread adoption".

CAESAR submissions specify a *family* of (one or more) authenticated ciphers the members of which may vary in external parameters.

CAESAR submissions follow an API specification consisting of:

$$c = \text{Encrypt}(p, k, \text{optional } d, \text{optional } mn_s, \text{optional } mn_p)$$

For ciphertext $c$, plaintext $p$, secret key $k$, optional associated data $d$, optional secret message number $mn_s$ and optional public message number $mn_p$.

See figure 4 for the different security purposes and restrictions of the inputs.

|  | **Integrity** | **Confidentiality** | **Length** | **Optional** | **May impose single-use requirements** |
|---|---|---|---|---|---|
| *Plaintext* | Yes | Yes | Variable | No | No |
| *Associated Data* | Yes | No | Variable | Yes | No |
| *Secret Message Number* | Yes | Yes | Fixed | Yes | Yes |
| *Public Message Number* | Yes | No | Fixed | Yes | Yes |
| *Key* | N/A | N/A | Fixed | No | N/A |

**Fig. 4**: *Security purposes and restrictions of inputs* [17]

CAESAR submission ciphers are expected to maintain security regardless of the user's choice of message numbers with the exception of reuse of a single (*secret message number*, *public message number*) pair for two encryptions with the same key.

## 3    Keccak

Keccak [18] is a family of *sponge functions* (see section 3.1) a subset of which has been standardized by NIST as the SHA-3 cryptographic hash function [19]. Keccak uses one of seven permutations denoted as $Keccak_f[b]$ (where *permutation width* b ∈ {25,50,100,200,400,800,1600}) as primitive in the sponge construction, with $Keccak_f[1600]$ being the permutation of choice for SHA-3.

Each of these permutations consists of the iterative application of a simple round function (similar to block ciphers but without key scheduling) with operations limited to bitwise XOR, AND, NOT and rotations.

The description in this section draws upon those outlined in [18,22,23].

### 3.1 The Sponge Construction

A *sponge construction* or *sponge function* [20] is a type of algorithm with a finite internal state consuming an arbitrary-length input bitstream and producing an output bitstream of an arbitrary desired length. Sponge functions can serve to implement various cryptographic primitives such as cryptographic hash functions, MACs, stream ciphers, PRNGs and AE schemes due to its arbitrarily long input and output sizes.

Sponge constructions are iterated constructions operating on an internal state $S$ of *width* $b$ bits (where $b = (r + c)$ with $r$ denoting the *bitrate* and $c$ denoting the *capacity*) by applying the fixed-length permutation function $f$ to it as illustrated in figure 5.

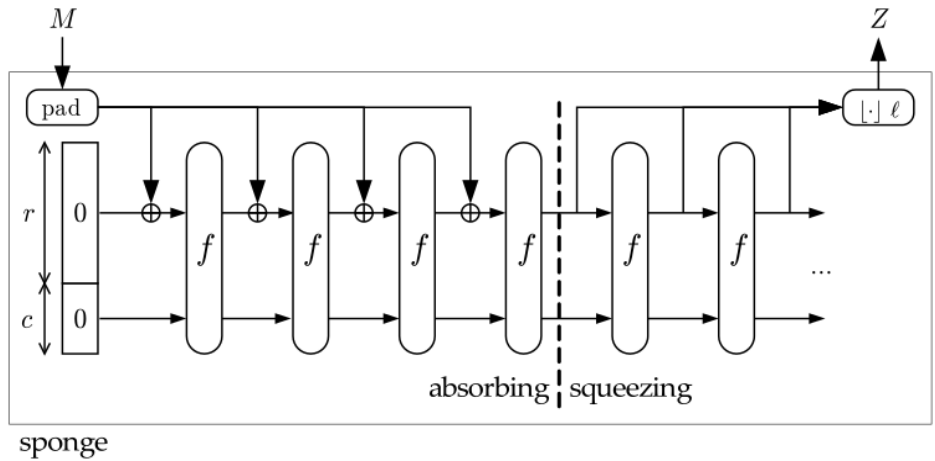

**Fig. 5**: *Sponge construction* [20]

The sponge function starts by padding the input string to a length that is a multiple of $r$ using a reversible padding rule before cutting it into $n$ blocks of $r$ bits. Next $b$ bits of state $S$ are initialized to zero and the sponge construction proceeds in two phases:

- *Absorbing*: In the absorbing phase every $r$-bit input block is XORed into the first $r$ bits of the state $S$ followed by application of the permutation function $f$ to $S$ resulting in a new state $S'$. After all input blocks have been processed in this fashion (ie. the padded input has been fully absorbed) the sponge construction switches to the squeezing phase.

- *Squeezing*: In the squeezing phase the user chooses the number of $r$-bit sized output blocks. For every block the first $r$ bits of the state $S$ are returned followed by application of the permutation function $f$ to $S$.

Note that the last $c$ bits (the *capacity*) of the state are never directly affected by the input blocks during the absorbing phase and are never output during the squeezing phase.

An example sponge construction API would look as follows:

$$Z = Sponge[f, pad, r](M, l)$$

For input message $M$ and requested output length $l$.

### 3.2 The $Keccak_f$ permutation function

There are 7 $Keccak_f$ [18] permutation functions denoted as $Keccak_f[b]$ where $b = 25 * 2^l$ and $0 \leq l \leq 6, l \in \mathbb{Z}$. One of the advantages of a permutation over traditional usage of block ciphers is that there is no need for a key schedule nor for an inverse.

The $Keccak_f$ function operates over an internal state $S$ which is represented as an array of 5x5 lanes each of length $w = 2^l$. The state can be divided in various parts (depending on axis or axes along which we select our bits) each with their own name as shown in figures 6a-c where the $x$ and $y$ axes represent a selection among the 5x5 lanes and the $z$ axis a $w$-bit sized lane.

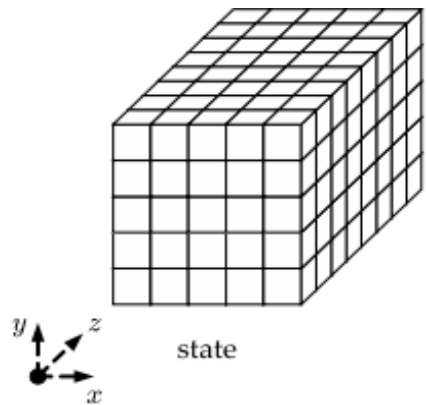

**Fig. 6a**: $Keccak_f$ *full state* [21]

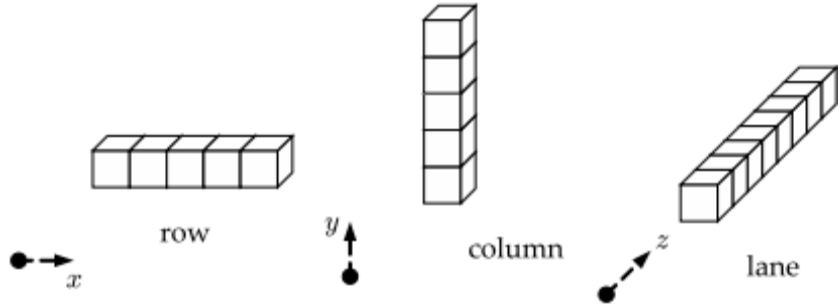

**Fig. 6b**: $Keccak_f$ state row, column and lane [21]

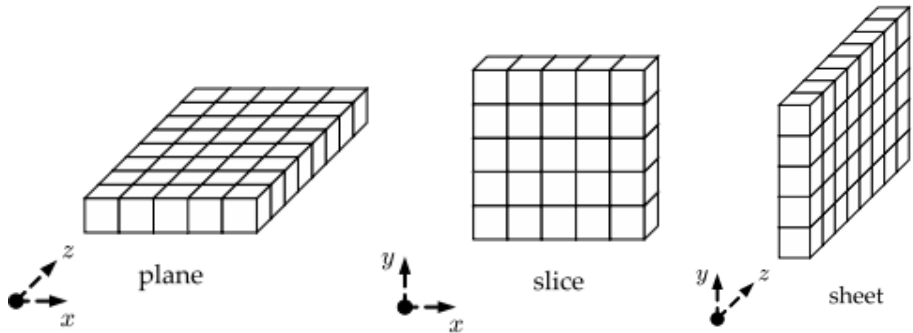

**Fig. 6c**: $Keccak_f$ state plane, slice and sheet [21]

The $Keccak_f[b]$ function consists of the iterated application of a *round function* with the permutation-width dependent number of rounds $n_r$ given by $n_r = 12 + 2l$ where $w = 2^l$ as illustrated in figure 7.

```
Keccak-f[b](S):
    for i ∈ {0, .., (n_r − 1)}:
        S = Round[b](S, RC_i)
    Return S
```

**Fig. 7**: $Keccak_f$ permutation *pseudo-code* [22]

The Keccak round function, as illustrated in figure 8, consists of the application of five steps $\theta, \rho, \pi, \chi, \iota$ to the state $S$ and a round constant $RC_i$ (where $i$ is the current round index) as defined in the round constant table in figure 9.

```
Round[b](A, RC):
  //θ-step for bit diffusion
  for x ∈ {0,..,4}:
    C[x] = A[x,0] xor A[x,1] xor A[x,2] xor A[x,3] xor A[x,4]

  for x ∈ {0,..,4}:
    D[x] = C[x-1] xor rot(C[x+1], 1)

  for (x, y) ∈ {{0,..,4}x{0,..,4}}:
    A[x,y] = A[x,y] xor D[x]

  //ρ-step for inter-slice diffusion and
  //π-step for disturbing x,y alignment through lane-transposition
  for (x, y) ∈ {{0,..,4}x{0,..,4}}:
    B[y, 2x + 3y] = rot(A[x,y], r[x,y])

  //χ-step for non-linear mapping
  for (x, y) ∈ {{0,..,4}x{0,..,4}}:
    A[x,y] = B[x,y] xor ((not B[x+1,y]) and B[x+2,y])

  //ι-step to break symmetry
  A[0,0] = A[0,0] xor RC

  Return A
```

**Fig. 8**: Round *pseudo-code* [22]

In figure 8 $A$ denotes the complete permutation state array (with $A[x, y]$ denoting a particular lane) whereas $B[x, y], C[x]$ and $D[x]$ are intermediate variables. The constants $r[x, y]$ are rotation offsets as listed in the table in figure 10. The $rot$ operation is a bitwise cyclic right-ward shift operation moving a bit at position $i$ to position $(i + r) \bmod w$.

| | |  | | |
|---|---|---|---|---|
| RC[ 0] | 0x0000000000000001 | RC[12] | 0x000000008000808B |
| RC[ 1] | 0x0000000000008082 | RC[13] | 0x800000000000008B |
| RC[ 2] | 0x800000000000808A | RC[14] | 0x8000000000008089 |
| RC[ 3] | 0x8000000080008000 | RC[15] | 0x8000000000008003 |
| RC[ 4] | 0x000000000000808B | RC[16] | 0x8000000000008002 |
| RC[ 5] | 0x0000000080000001 | RC[17] | 0x8000000000000080 |
| RC[ 6] | 0x8000000080008081 | RC[18] | 0x000000000000800A |
| RC[ 7] | 0x8000000000008009 | RC[19] | 0x800000008000000A |
| RC[ 8] | 0x000000000000008A | RC[20] | 0x8000000080008081 |
| RC[ 9] | 0x0000000000000088 | RC[21] | 0x8000000000008080 |
| RC[10] | 0x0000000080008009 | RC[22] | 0x0000000080000001 |
| RC[11] | 0x000000008000000A | RC[23] | 0x8000000080008008 |

**Fig. 9**: *Round Constant table* [22]

Round Constants are given by $RC[i][0,0,2^j - 1] = rc[j + 7i]$ for $0 \leq j \leq l$ where $i$ is the round index and all other $RC[i][x, y, z]$ are zero. The values $rc[t] \in \text{GF}(2)$ are given by the LFSR $rc[t] = (x^t \bmod x^8 + x^6 + x^5 + x^4 + 1) \bmod x$ in $\text{GF}(2)[x]$ with period 255.

| | x=3 | x=4 | x=0 | x=1 | x=2 |
|---|---|---|---|---|---|
| y=2 | 25 | 39 | 3 | 10 | 43 |
| y=1 | 55 | 20 | 36 | 44 | 6 |
| y=0 | 28 | 27 | 0 | 1 | 62 |
| y=4 | 56 | 14 | 18 | 2 | 61 |
| y=3 | 21 | 8 | 41 | 45 | 15 |

**Fig. 10**: *Rotation Offset table* [22]

### 3.3 Keccak's *pad10*1* Padding

The (reversible) padding rule used in Keccak [18] pads a message $M$ to a sequence of $x$-bit blocks and is denoted by $M||pad[x](|M|)$. Keccak's multi-rate padding, denoted by *pad10*1*, appends a single bit 1 followed by the minimum number of bits 0 followed by a single bit 1 such that the length of the result is a multiple of the block-length.

### 3.4 The Keccak Sponge Function

When combining the above documented functions into a sponge construction we obtain the $Keccak[r,c]$ sponge function (where $r$ denotes bitrate and $c$ denotes capacity) as illustrated in figure 11. Note that the pseudo-code in figure 11 is limited to cases where the number of bits in a message $M$ is a multiple of 8 (ie. spans a whole number of bytes) and $r$ is a multiple of the lane size (as is the case for the SHA-3 parameters [19]. In the pseudo-code $S$ denotes the state as an array of lanes with padded message $P$ organized as an array of blocks $P_i$ each of which is organized as an array of lanes themselves.

```
Keccak[r,c](M):
  //Initialization and padding
  for (x, y) ∈ {{0,..,4}x{0,..,4}}:
    S[x,y] = 0
  P = M || 0x01 || 0x00 || … || 0x00
  P = P xor (0x00 || … || 0x00 || 0x80)

  //Absorbing phase
  for P_i ∈ P:
    for (x, y) such that x + 5 * y < r/w:
      S[x,y] = S[x,y] xor P_i[x + 5y]
    S = Keccak-f[r+c](S)

  //Squeezing phase
  Z = empty string
  while (output is requested):
    for (x, y) such that x + 5 * y < r/w:
      Z = Z || S[x,y]
    S = Keccak-f[r+c](S)

  Return Z
```

**Fig. 11**: $Keccak[r,c]$ *sponge function pseudo-code* [22]

## 4    Keyak

Keyak [24] is a permutation-based authenticated encryption scheme supporting sessions submitted to the 2nd round of the CAESAR competition. In this document when we mention Keyak we are referring to Keyak v2 unless explicitly stated otherwise.

Keyak operates in the so-called *Motorist* mode (see section 4.2) and uses *Keccak-p* (see section 4.3) as the underlying permutation and is free of inverses [29]. Keyak offers features such as in-place encryption, parallelizable encryption/decryption, incremental AEAD [29] simultaneous processing of plaintext and associated data, session support, intermediate authentication tags and the option to combine wrapping and unwrapping in the same session where upon unwrapping the plaintext is made available only if the tag is valid. Keyak consists of five named instances: *River Keyak, Lake Keyak, Sea Keyak, Ocean Keyak* and *Lunar Keyak* taking on specific parameter values as detailed in section 4.7.

The material in this section draws upon that outlined in [20, 24, 25, 26, 27, 28].

### 4.1    The Duplex Construction

A *duplex construction* [20] is a construction closely related to the *sponge construction* with an equivalent level of security. The duplex construction allows for the alternation of input and output blocks (as shown in figure 12) at the same rate as the sponge construction, analogous to full-duplex communication, which allows for efficient implementations of reseedable PRNGs and AE schemes requiring only one call to the permutation function $f$ per input block.

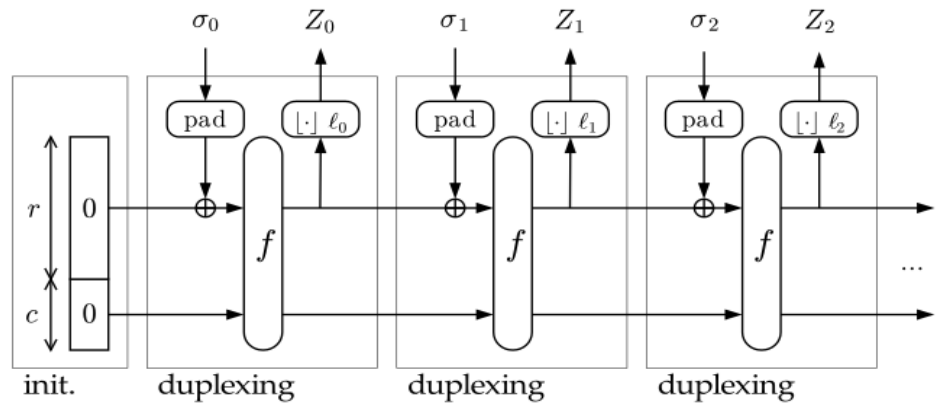

**Fig. 12**: *Duplex construction* [20]

The duplex construction, like the sponge construction, starts by padding the input string to a length that is a multiple of $r$ using a reversible padding rule before cutting it into $n$ blocks of $r$ bits. Next $b$ bits of state $S$ are initialized to zero. Unlike the sponge construction, which is stateless in between calls, the duplex construction results in an object which accepts calls taking an input string and returning an output string which depends on *all* input received so far. Such an object, called a *duplex object*, is denoted as $D$.

A duplex object $D$ has a $b$-bit internal state $S$ which is set to zero upon initialization. From then one can make calls to $D.duplexing(\sigma, l)$ (where $\sigma$ denotes an input string and $l$ denotes the requested number of bits) where the maximum number of bits one can request is the bitrate $r$ and the input string $\sigma$ has to be short enough such that after padding it results in a *single* $r$-bit block. This maximum length of $\sigma$ is called the *maximum duplex rate* and denoted as $\rho_{max}(pad, r)$ and is always smaller than the bitrate $r$.

Executing a $D.duplexing(\sigma, l)$ call will have the duplex object pad the input string $\sigma$ and XOR it to the first $r$ bits of the internal state $S$ after which it applies permutation function $f$ to the state and returns the first $l$ bits of the state as output. Calls where $\sigma$ is the empty string are referred to as *blank calls* while calls with $l = 0$ are referred to as *mute calls*.

An example duplex construction API would look as follows:

- *Initialization*: $D = Duplex[f, pad, r]$
- *Duplexing calls*: $D.duplexing(\sigma, l)$

### 4.2  Keyed Sponge Constructions

Keyed sponge constructions can be divided in two categories:

- *Outer-keyed Sponge constructions*: Bertoni et al. [26] introduced the keyed Sponge construction as an evaluation of the Sponge function over a concatenation of the key and the message, ie. $Sponge(K||M)$. This type of keyed Sponge is denoted as *outer-keyed*.

- *Inner-keyed Sponge constructions*: Chan et al. [27] introduced the *inner-keyed Sponge construction* which takes the regular sponge construction and uses $E_K^f$ as the Sponge permutation. Here $f$ is a permutation, $K$ the key and $E_K^f$ the *Even-Mansour construction* [28] which builds a $b$-bit block cipher from $b$-bit permutation using a $b$-bit key: $E_K^f = f(x \oplus K) \oplus K$.

### 4.3 Full-State Keyed and Duplex Sponge Constructions

In order to optimize the efficiency in Sponge-based authenticated encryption Mennink et al. [25] formalized the *Full-State Duplex Sponge* (FDS) construction which differs from the original duplex construction (see section 4.1) in that the key is explicitly used to initialize the state (making it a keyed construction) and the absorption phase is performed on the entire state which enforces explicit usage of the key. Mennink et al. [25] proved the increase in input block length from bitrate $r$ to full-state permutation width $b$ has no noticeable impact on the security of the generic construction while allowing for the injection of more bits per call to the underlying permutation and thus improving performance.

Note that in full-state keyed sponge constructions the usage of outer-keyed sponge constructions makes no longer any security difference from the usage of inner-keyed sponge constructions and both can be seen as special cases of the *Full-State Keyed Sponge* (FKS) construction.

#### 4.3.1 Full-State Keyed Sponge Constructions

The *Full-State Keyed Sponge* (FKS) constructions works by initializing the inner $k$ bits of the state $S$ to the key $K$ (where $k \leq c$) and the outer $(b - k)$ bits to zero. The message $M$ is padded to a bit-length that is a multiple of $b$ and absorbed in the usual sponge fashion. After absorption the squeezing phase outputs the outer $r$ bits of state $S$ in the usual sponge fashion until the request amount of $z$ bits is output as illustrated in figure 13.

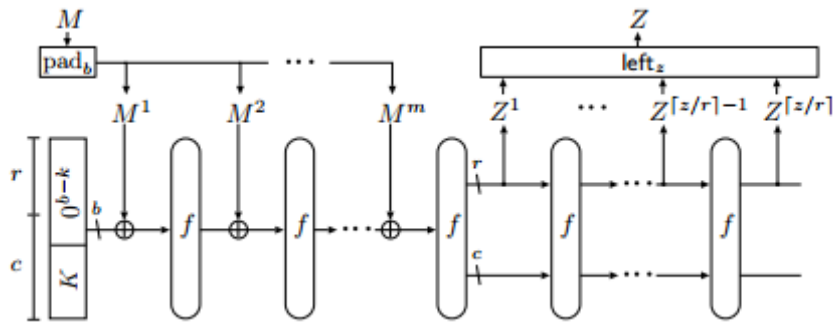

**Fig. 13**: *Full-State Keyed Sponge (FKS) construction* [25]

### 4.3.2 Full-State Duplex Constructions

The *Full-State Duplex Sponge* (FDS) constructions works by initializing the inner $k$ bits of the state $S$ to the key $K$ (where $k \leq c$) and the outer $(b - k)$ bits either zero (as per figure 14a). In the *Full-State Keyed Duplex* (FSKD) construction outlined in [24] and shown in figure 14b initialization consists of setting the inner $k$ bits of state $S$ to key $K$ and the outer $(b - k)$ bits to string $\sigma_0$ followed by application of the permutation $f$. Essentialy initialization of the latter is identical to initialization of the former followed by a single duplexing call with (unpadded) $M^1 = \sigma_0$. This results in duplex object $D$. Subsequent duplexing calls to $D$ can consume messages of up to $b$ bits while outputting string $Z$ as illustrated in figures 14a and 14b. Both constructions are identical save for the slight difference in initialization.

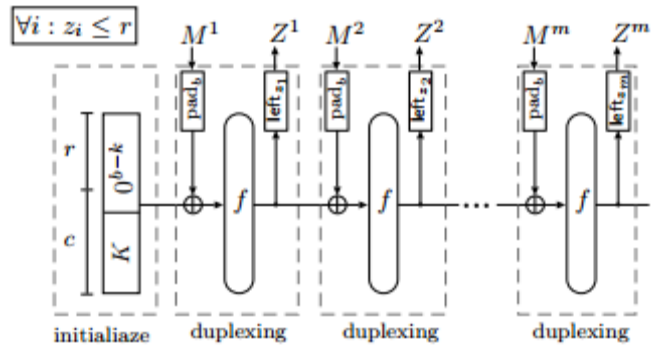

**Fig. 14a**: *Full-State Duplex Sponge (FDS) construction* as per [25]

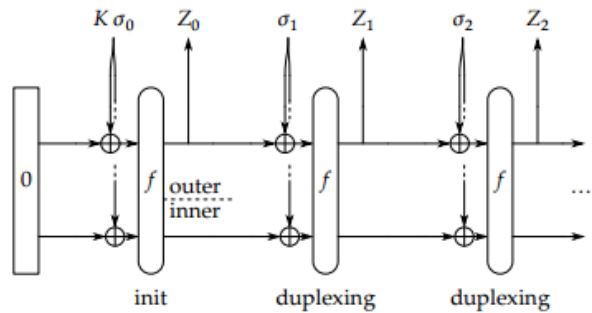

**Fig. 14b**: *Full-State Keyed Duplex (FSKD) construction* as per [24]

The operation of the Motorist mode can be expressed in calls to FSKD objects.

### 4.4 The Motorist mode

The Motorist mode is a construction supporting authenticated encryption of sequences of messages in sessions and serves as a replacement for *DuplexWrap* and *KeyakLines* used in Keyak v1. In a given session it processes messages and '*cryptograms*' where a message consists of a plaintext and possible associated data (which will be referred to as *metadata* from here on) and a cryptogram consists of a ciphertext, possible metadata and an authentication tag. Messages can consist of metadata alone and the corresponding cryptogram will not have any ciphertext.

- *Wrapping*: Each message is *wrapped* into a cryptogram by encrypting it into a ciphertext and computing an authentication tag over the full sequence of messages.

- *Unwrapping*: Each cryptograph is *unwrapped* by decrypting the ciphertext into plaintext, verifying the authentication tag and returning the plaintext if the tag is valid.

Within a session the tag of a given cryptogram authenticates the full sequence of messages sent/received since the start of that particular session. Starting a session requires a secret key (and possibly a nonce if the secret key is not unique for this session).

#### 4.4.1 Motorist's Duplex Instances

The Motorist mode is sponge-based and supports one or more duplex instances, in the form of so-called *Pistons* (see section 4.4.3.1) operating in parallel. It calls the duplexing function with input containing the key, nonce, plaintext and metadata bits and uses the output as tag or key stream bits.

The Motorist duplex instances differ from the original duplex construction (see section 4.1) in that they are *full-state keyed duplex* (FSKD) instances (see section 4.3.2) and accept input blocks as large (after padding) as the permutation width instead of only the bitrate width. The Motorist mode supports a parameterized degree of parallelism in the form of an array of so-called *Pistons* (see section 4.4.3.1) each of which is an FSKD object.

Motorist distributes the message (consisting of plaintext and metadata) over a set of different *Piston* duplex instances. In order to produce an authentication tag that depends on the full message and not only on the message bits supplied to a single duplex instance, Motorist performs dedicated processing (called a *knot*) at the end of each message which extracts chaining values from each duplex instance and after concatenating them injects them into all duplex instances resulting in the state of all instances depend on the full sequence of messages. Finally it extracts an authentication tag from a single duplex object which authenticates the full message sequence (ie. session).

#### 4.4.2 Motorist's Session Support

In order to start a session Motorist consumes a globally unique and secret string called the *Secret and Unique Value* (SUV) consisting of a key and a nonce (with the recommended order being the key coming first) which is injected into each duplex instance followed by the appending of a diversification string to diversify their states. A single Motorist session is sufficient for secure two-way communication but one must clearly indicate for every message who is the sender which can be done by including the sender's identifier in the message metadata. An alternative approach would be relying on a strict alternating convention. The SUV nonce requirement is not required for Motorist sessions which perform unwrapping functionality only.

#### 4.4.3 Motorist's Layers

Motorist is specified in three layers each of which handle a different aspect of its functionality each consuming input in terms of *byte streams* (that is, strings of bytes that can be read from and written to in sequential fashion) where a sequence of consecutive bytes from a stream is called a *fragment*.

##### 4.4.3.1 Piston Layer

The *Piston layer* is effectively an augmented *full-state keyed duplex* (FSKD) construction that maintains a $b$-bit state $S$ and applies permutation $f$ to it and has further parameters squeezing rate $R_s$ and absorbing rate $R_a$ where $R_s \leq R_a$. It handles basic functionality such as data injection, simultaneous encryption or decryption (if so desired), tag extraction and setting fragment offsets. The Piston state $S$ is initialized to zero and $s[i]$ denotes byte $i$ of state $S$.

An example Piston API would look as follows:

$$Piston[f, R_s, R_a]$$

Where $f$ denotes the underlying permutation function, $R_s$ the squeeze rate and $R_a$ the absorbing rate.

When properly used, ie. through an *Engine* (see section 4.4.3.2), the Piston constructs a full-width input block from supplied plaintext, metadata and fragment offsets encoding as illustrated in figures 15a and 15b where either:

- The block starts with a possible number of leading zeros (starting at index 0), followed by a plaintext fragment ending at index $R_s$, followed by a metadata fragment ending at index $R_a$, followed by the fragment offsets.

- In the absence of plaintext metadata starts at index 0 and runs up to index $R_a$.

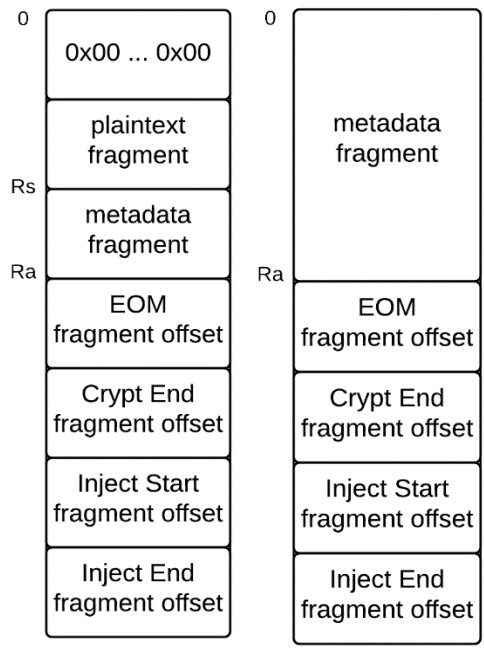

**Fig. 15a,b**: *Piston full-width input blocks with (a) and without (b) plaintext*

Piston input blocks have four fragment offsets:

- *EOM*: The *EOM* fragment offset functions both to denote the number of bytes in the next output block to be used as tag and to delimit messages by having a non-zero value if it is part of an input block that is the last of a message or a string injected collectively. In case no tag is requested at the end of such a collectively injected message or string it takes the value 255 (0xFF).

- *Crypt End*: The *Crypt End* fragment offset functions to denote the end of the plaintext fragment within the current input block. The start of this plaintext fragment is denoted by the EOM fragment offset of the previous input block where a value of 255 (0xFF) means the plaintext fragment starts at index 0.

- *Inject Start*: The *Inject Start* fragment offset functions to denote the start of the metadata fragment within the current input block. If there is also a plaintext fragment within the current input block then the metadata fragment starts at $Inject\ Start = R_s$ else it starts at $Inject\ Start = 0$.

- *Inject End*: The *Inject End* fragment offset functions to denote the end of the metadata fragment within the current input block.

After the input block has been absorbed into the state and permutation function $f$ has been applied the outer part of the state is used as follows:

- An arbitrary number of bytes used as tag, starting at index 0

- An arbitrary number of bytes used as keystream, starting after the possible tag

The Piston has four functions (replacing the traditional single *Duplexing* function of a Duplex object), illustrated in figures 16a-d, achieving the above functionality:

- $Piston.Crypt(I, O, \omega, unwrapFlag)$: This function, as illustrated in figure 16a, supports the combined encryption of plaintext (or decryption of ciphertext) and absorbing of the corresponding ciphertext (or plaintext) into the outer part of the state as long as input is available followed by updating the end of the plaintext fragment in the appropriate fragment offset. The parameters are as follows:

    - $I$: Input bytestream (either plaintext or ciphertext)
    - $O$: Output bytestream where result will be written to
    - $\omega$: Index in the state from where the plaintext fragment must be injected. The fragment will end at index $R_s$ or earlier (if input is exhausted). The end of the plaintext fragment is set in offset $Crypt\ End$.
    - $unwrapFlag$: Indicates whether we are encrypting (False) or decrypting (True)

```
Crypt(I, O, ω, unwrapFlag):
  while (hasMore(I) and (ω < R_s))
  {
    x = I.get()
    O.put(state[ω] ^ x)
    state[ω] = x if (unwrapFlag) else (state[ω] ^ x)
    ω++
  }
  state[cryptEnd] ^= ω
  Return
```

**Fig. 16a**: $Piston.Crypt()$ *pseudo-code* [24]

- $Piston.Inject(X, cryptingFlag)$: This function, illustrated in figure 16b, injects metadata taken from input bytestream $X$ starting from index $R_s$ if $cryptingFlag$ is set (indicating whether the current input block already has a plaintext fragment) or from index 0 if it is not. The metadata fragment will end at index $R_a$ or earlier (if input is exhausted). The start of the metadata fragment is set in offset $Inject\ Start$.

```
Inject(X, cryptingFlag):
  ω = R_s if (cryptingFlag) else 0
  state[injectStart] ^= ω
  while (hasMore(X) and (ω < R_a))
  {
    state[ω] ^= X.get()
    ω++
  }
  state[injectEnd] ^= ω
  Return
```

**Fig. 16b**: $Piston.Inject()$ *pseudo-code* [24]

- $Piston.Spark(eomFlag, l)$: This function, as illustrated in figure 16c, applies the underlying permutation $f$ to the state $S$. Before doing so it sets in the $EOM$ fragment offset whether this is the last input block of a message (or of a string collectively injected) as indicated by $eomFlag$. If this is the case $EOM$ is set to the number $l$ of bytes of the state $S$ (after application of $f$) which are reserved as tag.

```
Spark(eomFlag, l):
  if (eomFlag):
    state[EOM] ^= 255 if (l=0) else l
  else
    state[EOM] ^= 0
  state = f(state)
  Return
```

**Fig. 16c**: $Piston.Spark()$ *pseudo-code* [24]

- $Piston.GetTag(T, l)$: This function writes the first $l$ bytes of the state $S$ to output byte stream $T$ to be used as a tag or chaining value.

> **GetTag(T, l):**
> **assert** $(l \leq R_s)$
> T.put(state[i]) **for** $i \in \{0, .., l-1\}$
> **Return**

**Fig. 16d**: $Piston.GetTag()$ *pseudo-code* [24]

#### 4.4.3.2 Engine Layer

The *Engine layer* controls and relies on an array of Π Piston objects operating in parallel. For every Piston the Engine maintains an attribute $E_t$ denoting how much output was used as tag or chaining value in order to pass this to $Piston.Crypt()$ and avoid the reuse of the bits as key stream.

An example Engine API would look as follows:

$$Engine[\Pi, Pistons]$$

Where Π denotes the number of parallel Pistons and $Pistons$ denotes an array of Π Piston objects.

The Engine also maintains a state machine in the *Phase* attribute in order to direct the function call sequence in order to maintain consistency. The *Phase* indicates how the Π input blocks are being constructed in the Π Piston objects with the following phases being available:

- $Fresh$: The input blocks are empty

- $Crypted$: The input blocks have a plaintext fragment and more plaintext is coming

- $EndOfCrypt$: The input blocks have a plaintext fragment and no more plaintext is coming

- $EndOfMessage$: The input blocks have their fragments ready and the message has been fully injected

The engine has five functions, illustrated in figures 17a-e, achieving the above functionality:

- *Engine. Spark(eomFlag, l)*: This function centralizes application of permutation $f$ and calls the *Piston. Spark(eomFlag, $l_i$)* function for each of the $\Pi$ Pistons, where $l_i \in l$ with $l$ being a vector. After sparking the Pistons the $E_t$ value is set to $l$.

> **Spark**(eomFlag, l):
>   Pistons[i].Spark(eomFlag, *l*[i]) **for** $i \in \{0,..,\Pi - 1\}$
>   $E_t = l$
>   **Return**

**Fig. 17a**: *Engine. Spark*() *pseudo-code* [24]

- *Engine. Crypt(I, O, unwrapFlag)*: This function dispatches input $I$ to the Pistons but requires the Engine to be in the *Fresh* phase. It subsequently calls *Piston. Crypt(I, O, $E_t[i]$, unwrapFlag)* on all of them where $i$ is the Piston index. If $I$ has not been exhausted (ie. there is more input stream) then the phase is set to *Crypted* else it is set to *EndOfCrypt*.

Note that each Piston takes a fragment of $I$ and so the Pistons process up to $\Pi R_s$ bytes.

> **Crypt**(I, O, unwrapFlag):
>   **assert** (phase = *fresh*)
>   Pistons[i].Crypt(I, O, $E_t[i]$, unwrapFlag) **for** $i \in \{0,..,\Pi - 1\}$
>   phase = *crypted* **if** (hasMore(I)) **else** *endOfCrypt*
>   **Return**

**Fig. 17b**: *Engine. Crypt*() *pseudo-code* [24]

- *Engine. Inject(A)*: This function dispatches metadata $A$ to the Pistons (provided the phase is not *EndOfMessage*) by calling *Piston. Inject(A, (Phase $\in$ {Crypted, EndOfCrypt}))* on each of them. If both input and metadata streams are exhausted the phase is set to *EndOfMessage* and application of $f$ is delayed until a call to *Engine. GetTags*() is made. Otherwise *Engine. Spark(False, $0^\Pi$)* is called and the phase is reset to *Fresh*.

Note that each Piston takes a fragment of $A$ and so the Pistons process up to $\Pi(R_a - R_s)$ bytes if *Engine. Crypt*() was called before and $\Pi R_a$ bytes otherwise.

```
Inject(A):
  assert (phase ∈ {fresh, crypted, endOfCrypt})
  cryptingFlag = (phase ∈ {crypted, endOfCrypt})

  Pistons[i].Inject(A, cryptingFlag) for i ∈ {0,..,Π − 1}
  if ((phase = crypted) or (hasMore(A))):
  {
    Spark(false, {0x00}^Π)
    phase = fresh
  }
  else:
    phase = endOfMessage
  Return
```

**Fig. 17c**: $Engine.Inject()$ *pseudo-code* [24]

- $Engine.GetTags(T, l)$: This function, which can only be called if the phase is $EndOfMessage$, calls $Engine.Spark(True, l)$ and collects the corresponding tags in output stream $T$ from all Pistons by calling $Piston.GetTag(T, l[i])$ on them where $i$ is the Piston index. It then sets the phase to $Fresh$ again.

```
GetTags(T, l):
  assert (phase = endOfMessage)
  Spark(true, l)
  Pistons[i].GetTag(T, l[i]) for i ∈ {0,..,Π − 1}
  phase = fresh
  Return
```

**Fig. 17d**: $Engine.GetTags()$ *pseudo-code* [24]

- $Engine.InjectCollective(X, diversifyFlag)$: This function, which can only be called if the phase is $Fresh$, injects the same metadata $X$ to all Piston objects by calling $Piston.Inject(X_t[i], False)$ on them and is used to inject the $SUV$ and chaining values. When $diversifyFlag$ is set (as set when injecting the $SUV$) it appends to $X$ two bytes:

  o One byte encoding the degree of parallelism Π (for domain separation between instances with a different number of Piston objects)

  o One byte encoding the index of the Piston object (for domain separation between Piston objects, in particular to avoid identical keystreams)

After the whole stream $X$ is processed the phase is set to $EndOfMessage$.

```
InjectCollective(X, diversifyFlag):
  assert (phase = fresh)
  X_t = {byteStream()}^Π

  while(hasMore(X))
  {
   x = X.get()
   X_t[i].put(x) for i ∈ {0,..,Π − 1}
  }

  if(diversifyFlag):
   (X_t[i].put(Π); X_t[i].put(i)) for i ∈ {0,..,Π − 1}

  X_t[i].seek(0,0) for i ∈ {0,..,Π − 1}

  while(hasMore(X_t[0]))
  {
   Pistons[i].Inject(X_t[i], 0) for i ∈ {0,..,Π − 1}
   if(hasMore(X_t[0])):
     Spark(False, {0x00}^Π)
  }
  phase = endOfMessage
  Return
```

**Fig. 17e**: $Engine.InjectCollective()$ *pseudo-code* [24]

#### 4.4.3.3 Motorist Layer

The *Motorist layer* controls an Engine object which controls a number Π of Piston objects. The Motorist layer has a parameter $W$ (typically 32 or 64) denoting the *alignment unit* which ensures fragment start offsets, tag lengths, chaining values and fragments (except upon stream exhaustion) are a multiple of $W$ allowing for multi-byte chunk data manipulation. The Motorist layer finally also has parameters which determine security strength: capacity $c$ and tag length $\tau$. From the latter the following is determined:

- $R_s$: The squeezing rate is the largest multiple of $W$ such that at least $c$ bits (for the inner part) or 32 bits (for the fragment offsets) of the state are never used as output.

- $R_a$: The absorbing rate is the largest multiple of $W$ such that at least 32 bits at the end of the state are reserved for absorbing the fragment offsets.

- $c'$: The chaining value length is the smallest multiple of $W$ greater than or equal to capacity $c$.

An example Motorist API would look as follows:

$$Motorist[f, \Pi, W, c, \tau]$$

The Motorist maintains a *Phase* attribute with the following possible values:

- *Ready*: The Motorist object has been initialized and no input has been given yet.

- *Riding*: The Motorist object processed the $SUV$ and can no wrap or unwrap. This remains the phase until an error occurs.

- *Failed*: The Motorist object received an incorrect tag.

The motorist has four functions, illustrated in figures 18a-d, achieving the above functionality:

- $Motorist.StartEngine(SUV, tagFlag, T, unwrapFlag, forgetFlag)$:
  This function, which can only be called if the phase is *Ready*, starts a session with a given $SUV$ read from the $SUV$ byte stream. It is collectively injected with the $diversifyFlag$ set for domain separation by calling $Engine.InjectCollective(SUV, True)$. If the parameter $forgetFlag$ is set a *knot* is needed and $Motorist.MakeKnot()$ is called after which $Motorist.HandleTag(tagFlag, T, unwrapFlag)$ is called. This last step supports generation of verification of a tag by means of $tagFlag$. If $unwrapFlag$ is set it verifies tag $T$ else it returns a tag in $T$. If this succeeds then the phase is set to *Riding*.

```
StartEngine(SUV, tagFlag, T, unwrapFlag, forgetFlag):
  assert (phase = ready)
  Engine.InjectCollective(SUV, true)

  if (forgetFlag):
    MakeKnot()

  r = HandleTag(tagFlag, T, unwrapFlag)
  if (r):
    phase = riding

  Return r
```

**Fig. 18a**: $Motorist.StartEngine()$ *pseudo-code* [24]

- $Motorist.Wrap(I, O, A, T, unwrapFlag, forgetFlag)$: This function, which can only be called if the phase is *Riding*, unwraps cryptograms or wraps messages (depending on if $unwrapFlag$ is set or not, respectively). The function starts by processing input $I$ and metadata $A$.

  If there is no available input or metadata it sets the right Engine phase by calling $Engine.Inject(A)$. As long as there is available input $I$ it calls $Engine.Crypt(I, O, unwrapFlag)$, where $O$ denotes the output stream, fol-lowed by a call $Engine.Inject(A)$ until there is no more available input. Next, as long as there is available metadata $A$ it calls $Engine.Inject(A)$ until there is no more available metadata. If either $\Pi > 1$ or $forgetFlag$ is set a call to $Motorist.MakeKnot()$ is made. Finally a call to $Motorist.HandleTag(True, T, unwrapFlag)$ is made and only if it succeeds does the $Motorist.Wrap$ call succeed and produce output in output stream $O$. A failing call to $Motorist.HandleTag$ results in a cleared output stream $O$.

This function can be called to either *wrap* or *unwrap* which is done as follows:

- o *Wrapping* : The function is to be called with $unwrapFlag = False$, a plaintext input stream $I$ and metadata stream $A$, an output stream $O$ for receiving the ciphertext and tag stream $T$ for receiving the tag as well as $forgetFlag$.

- o *Unwrapping*: The function is to be called with $unwrapFlag = True$, a ciphertext input stream $I$, metadata stream $A$, and tag stream $T$ and an output stream $O$ for receiving the plaintext as well as $forgetFlag$. The function returns $True$ *iff* the tag is correct and false otherwise. The function clears $O$ if the tag is incorrect.

```
Wrap(I,O,A,T, unwrapFlag, forgetFlag):
assert (phase = riding)

Engine.Inject(A) if (hasMore(I) and not (hasMore(A)))

while(hasMore(I)):
{
  Engine.Crypt(I, O, unwrapFlag)
  Engine.Inject(A)
}

while(hasMore(A)):
  Engine.Inject(A)

MakeKnot() if ((Π > 1) or (forgetFlag))
r = HandleTag(true, T, unwrapFlag)
O.erase() if not(r)

Return r
```

**Fig. 18b**: $Motorist.Wrap()$ *pseudo-code* [24]

- $Motorist.MakeKnot()$: This function, which is only used internally by the Motorist, is used to either make a tag depend on the state of $\Pi > 1$ Pistons or, when $\Pi = 1$, achieve forgetting. The function starts by retrieving a $c'$-bit chaining values from every Piston object by calling $Engine.GetTags(T', \left[\frac{c'}{8}\right]^\Pi)$, where $T'$ is an initially empty local bytestream, and concatenates them into a $\Pi c'$-bit string which is collectively injected into all Piston objects by calling $Engine.InjectCollective(T', False)$.

```
MakeKnot():
  T' = byteStream()
  Engine.GetTags(T', {c'/8}^Π)
  T'.seek(0,0)
  Engine.InjectCollective(T', false)
  Return
```

**Fig. 18c**: $Motorist.MakeKnot()$ *pseudo-code* [24]

- $Motorist.HandleTag(tagFlag, T, unwrapFlag)$: This function, which is only used internally by the Motorist, starts out with an initially empty byte stream $T'$ and if $tagFlag$ is set proceeds to obtain tags using $Engine.GetTags(T', \left[\frac{\tau}{8}, 0^{\Pi-1}\right])$ and followed by either copying $T'$ to $T$ (if $unwrapFlag = False$) to obtain a tag during wrapping or (if $unwrapFlag = True$) checking whether $T' \neq T$ and if so setting phase to $Failed$ in order to indicate that the resulting tag and desired tag did not match during unwrapping. If $tagFlag$ is not set it simply calls $Engine.GetTags(T', 0^\Pi)$ to move the engine phase along.

```
HandleTag(tagFlag, T, unwrapFlag):
  T' = byteStream()
  if not(tagFlag):
    Engine.GetTags(T', {0x00}^Π)
  else:
    l = {0x00}^Π;  l[0] = τ/8
    Engine.GetTags(T', l)
    if not(unwrapFlag): (T = T')
    else if not(T' = T):
      phase = failed;
      Return false
  Return true
```

**Fig. 18d**: $Motorist.HandleTag()$ *pseudo-code* [24]

After starting a session with $Motorist.StartEngine()$ the Motorist object can receive an arbitrary number of calls to $Motorist.Wrap()$. The nonce requirement of the $SUV$ holds at session level (within a session messages have no explicit number of nonce but must be processed in-order for tag verification). Both communicating parties must use synchronized values for the $tagFlag$ and $forgetFlag$ parameters.

### 4.5 The $Keccak_p$ permutations

The $Keccak_p$ permutations are derived from the $Keccak_f$ permutations (see section 3.2) and have a tweakable number of rounds. A $Keccak_p$ permutation is defined as $Keccak_p[b, n_r]$ where $b$ is the permutation width (where $b = 25 * 2^l$ and $0 \leq l \leq 6, l \in \mathbb{Z}$) and $n_r$ the number of rounds. In short $Keccak_p[b, n_r]$ consists of the application of the last $n_r$ rounds of $Keccak_f[b]$. In the case that $n_r = 12 + 2l$ then $Keccak_p[b, n_r] = Keccak_f[b]$.

### 4.6 The Key Pack

Keyak keys are encoded in so-called *Key Packs* which serve to encode secret keys as prefix of an $SUV$. Key packs make use of the $pad10*[r](|M|)$ padding rule which returns a bitstring $10^q$ where $q = (|M| - 1) \mod r$. If $r$ is a multiple of 8 and $M$ a sequence of bytes the padding rule returns the bytestring $0x01\ 0x00^{\frac{q-7}{8}}$. Given a key $K$ a key pack of $l$ bytes is defined as:

$$KeyPack(K, l) = enc_8(l) || K || pad10*[l - 8](|K|)$$

Where $l < 256$ indicates the full length of the key pack in bytes and key $K$ is limited to $8(l - 1) - 1$ bits resulting in the key pack illustrated in figure 19.

| length field (1) | key (k) | padding (l-(k+1)) |
|---|---|---|
| 0x12 | 0x01 0x23 .. 0xEF | 0x01 0x00 .. 0x00 |

**Fig. 19**: $KeyPack(K, 18)$ with 64-bit key $K$.

### 4.7 The Keyak Encryption Scheme

Combining the above documented functionality gives us the *Keyak* encryption scheme as an instantiation of the *Motorist* mode with a $Keccak_p$ instance as permutation. A Keyak instance is defined as follows:

$$Keyak[b, n_r, \Pi, c, \tau] = Motorist[Keccak_p[b, n_r], \Pi, W, c, \tau]$$

Where $W = max(\frac{b}{25}, 8)$.

The *SUV* is given as:

$$SUV = KeyPack(K, l_k) || N$$

Where $l_k = \frac{W}{8} \left\lceil \frac{c+9}{W} \right\rceil$ and no limitation on the length of nonce $N$.

#### 4.7.1 Keyak Named Instances

There are five named *Keyak* instances taking on various specific parameters and being suitable for various different optimizations. All instances have the following parameters: $n_r = 12, c = 256, \tau = 128$ (specifying the usage of 12 rounds, a capacity of 256-bits and a tag size of 128-bits respectively). The instances are, in order of increasing state size:

- *River Keyak*: $b = 800, \Pi = 1$

- *Lake Keyak*: $b = 1600, \Pi = 1$ (primary recommendation)

- *Sea Keyak*: $b = 1600, \Pi = 2$

- *Ocean Keyak*: $b = 1600, \Pi = 4$

- *Lunar Keyak*: $b = 1600, \Pi = 8$

For *River Keyak* $W = 32$ and key pack length $l_k = 36\ bytes$ while for the other instances $W = 64, l_k = 40\ bytes$. All instances take a (variable length) public message number or nonce $N$ but no private message number. If $N$ is to be fixed length it is proposed to be 58 bytes for *River Keyak* and 150 bytes for other instances. All instances produce a $\tau = 128$-bit tag (or MAC) which can be truncated if desired by the user. If not truncated the gap between plaintext and ciphertext length is 128 bits. Key sizes are variable with key size $128 \leq k \leq l_k$ (where the maximum is at least 256 bits).